\documentclass[preprint,showpacs,preprintnumbers,amsmath,amssymb,showkeys]{revtex4}

\usepackage{graphicx}% Include figure files
\usepackage{epsfig}% Include figure files
\usepackage{dcolumn}% Align table columns on decimal point
\usepackage{bm}% bold math

\begin{document}

\title{Properties of the doped spin 3/2 Mott insulator near half filling}
\author{Stellan \"Ostlund }  
\email{ostlund@fy.chalmers.se}
\affiliation{
Chalmers Technical University/Gothenburg University \\ Gothenburg 41296, Sweden \\
}

\author{ T. H. Hansson and A. Karlhede}
\affiliation{
Department of Physics, Stockholm University, AlbaNova University Center \\ SE-106 91 Stockholm, Sweden
}
\email{hansson@physto.se, ak@physto.se}
\newcommand{\cdop}[2]{c^{\dagger}_{#1,#2} }           % c creation operator
\newcommand{\cop}[2]{c^{\phdagger}_{#1,#2 }}
\newcommand{\gt}{>}
\newcommand{\ket}[1]{{ | \;  { #1 }  \; \rangle}}
\newcommand{\Deltasd}[1]{ { \Delta^{\dagger} }_{#1} }
\newcommand{\chat}[2]{{\hat{c}}^{\phdagger}_{#1,#2}}           % c annihilation operator
\newcommand{\chatd}[2]{{\hat{c}}^{\dagger}_{#1,#2}}            % c creation operator
\newcommand{\bra}[1]{{ \langle \; { #1 } \; | }}
\newcommand{\Q}[1]{Q^{\phdagger}_{#1}}
\newcommand{\Qd}[1]{Q^{\dagger}_{#1}}
\newcommand{\phdagger}{\phantom{\dagger}}
\newcommand{\subs}[2]{ {#1}_{#2} }
\newcommand{\sups}[2]{ {#1}^{#2} }
\newcommand{\expectation}[1]{\langle \;  #1 \; \rangle}
\newcommand{\e}[1]{\langle \;  #1 \; \rangle}
\newcommand{\nh}{\hat{n}}
\newcommand{\Deltah}[1]{{\hat{\Delta}_{#1}}}
\newcommand{\rp}{{r'}}
\newcommand{\hopi}[1]{\hat{h}_{#1} }
\newcommand{\hop}{{h}}
\newcommand{\half}{\frac{1}{2}}
\newcommand{\Deltat}[1]{{\tilde{\Delta}_{#1}}}
\newcommand{\Deltahd}[1]{{\hat{\Delta}_{#1}}^{\dagger}}
\newcommand{\dmu}{{\delta_\mu}}
\newcommand{\Deltatd}[1]{{\tilde{\Delta}_{#1}}^{\dagger}}

\date{\today}% It is always \today, today,
             %  but any date may be explicitly specified

\begin{abstract}

We develop an exact generalized Bogoliubov transformation for the spin $
3/2$ Hubbard model with large anti-Hunds rule coupling near half filling. 
Since our transformation is unitary, we can thereafter employ standard approximate 
mean field theory methods in the full Hilbert space to analyze the doped Mott insulator,
 in contrast to a
conventional approach  based on truncated Hilbert
spaces complemented with hard core constraints. 
The ground state at exactly half filling is an insulating (Mott) singlet,
and according to our  analysis  a non-Fermi liquid order parameter $ \Delta $ 
usually associated with extended 
s-wave superconductivity, will appear self-consistently 
as soon as a finite density $ n $  holes are introduced. 
The non-Fermi liquid behavior
is a consequence of the nonlinear nature of the unitary transformation
mapping the Mott singlet state to a Fock vacuum which introduces anomalous terms such as  $\Delta n$ in the effective Lagrangian.
Our analysis uses an approach that generalizes readily to
multi-band Hubbard models and could provide a mechanism
whereby a non-Fermi liquid order parameter
proportional to density is developed in Mott insulators
with locally entangled ground states. 
For more complicated systems, such an
order parameter could coexist naturally with a variety of other order parameters.

\end{abstract}

\pacs{Valid PACS appear here}
\maketitle

\section{Introduction}

A  Mott transition 
is expected to occur when the overlap between atomic orbitals
in an insulator becomes large enough for the hopping energy to overcome
the energy associated with charge fluctuations.
To understand this transition has turned out to be a very difficult problem that 
has been attacked by a variety of means, with dynamic mean field theory being
an important recent contribution
\cite{revmod68,science2364,PRL05361,condmat0306312,p134651,PRB07619,condmat0401041}.
In the present work we will develop an alternative approach, more in the
spirit of BCS theory, which is appropriate for understanding doped Mott insulators
where the parent state has an even number of electrons, and thus integer spin, 
at each state. Such models have been considered for instance in two-band Hubbard
models in the context of ruthenate alloys \cite{condmat0401223},
and multi-band Hubbard models in the theory of doped $ C_{60} $ \cite{granath1,granath2,PRL05361,
science2364,PRB07619}.
We will present evidence for non-Fermi liquid
behavior in these systems even for weak doping.  We shall 
focus on the simplest system to which our conclusions apply,
namely the spin $ 3/2 $ Hubbard model on a square lattice with anti-Hunds rule 
couplings \cite{p247_1,PRL8642,condmat0401223,condmat0401090,granath1,granath2}.

We define the local particle number
by $ n_{r} =   \sum_{s} \cdop{s}{r} \cop{s}{{r } } $ and the 
spin by
$ S_{r} =    \sum_{s} \cdop{s}{r} S_{s,s\prime}\cop{s\prime}{{r } } $, where
$ -\frac{3}{2} \le s \le \frac{3}{2} $ and $ S_{s,s\prime} $ are
the generators of spin $ 3/2 $ rotations.
Furthermore, we define the operator $ P^{\dagger} _{2,m}(r) $ which creates
an $ l=2, l_z = m  $ state  with two fermions \cite{PRL8642},
$  P^{\dagger}_{2,m}(r)  = \sum_{\alpha,\beta} \langle \frac{3}{2}\frac{3}{2}\alpha \beta|\frac{3}{2}\frac{3}{2}
;2,m\rangle \cdop{\alpha}{r} \cdop{\beta}{r} $. We also define the $ SU(2) $ invariant
$    P^2_r \equiv     \sum_m P^{\dagger}_{2,m}(r)  P_{2,m}(r) $. 
The Hamiltonian containing the maximal
number of onsite terms permitted by symmetry is
\begin{equation} \label{eq:hamiltonian}
H  =   - t  \sum_{r,\delta} \cdop{s}{r} \cop{s}{{r + \delta } }  +  
		\sum_{r}  (  U n_{r} ( n_{r} - 2 ) \; + \;    J   P^2_r ) \ ,
\end{equation}
where an arbitrary chemical potential is absorbed in $ U $.
Another term which can be considered is of course $ S^2_{r} $, but for spin $ 3/2 $
this obeys  
$    P^2_r = \frac{1}{3} S^2 - \frac{5}{2} n +
	 \frac{ 5}{4} n^2 $, 
so there is no loss of generality in the onsite term of our Hamiltonian aside from ignoring 
the possibility of an arbitrary  term proportional to $ n^3 $ . 
We will consider the case $  J \gt 2 U  \gg t  $, thus the singlet state is
heavily favored near $ n = 2 $. 

The single site spectrum consists of sixteen states: an empty site,
four equivalent spin $ 3/2 $ singly charged states, a singlet and
five spin $ 2 $ doubly charged sites, four spin $ 3/2 $ charge three
states and a charge four singlet.   If $ E_g(n) $ is the atomic ground
state for $n$ particles we find $ E_g(0) = E_g(2) = 0 $, $ E_g(1) = -U, E_g(3) = 5 J + 3 U $,
	$ E_g(4) = 10 J + 8 U $ and the quintet state has energy $ 2 J $. 
The lowest energy states obey $ E_g(n+1) + E_g(n-1) - 2 E_g(n) \ge U   \gg 0 $
so there is no tendency for superconducting pair formation from any of the 
local interactions.  

With our choice of parameters, standard arguments imply that the ground
state for $ n = 2 $ and small $ t $ should be a spin singlet. For
small doping near zero filling, the ground state will most likely be a
normal Fermi liquid.  This statement may be violated if $ U $ and $ J$
are sufficiently large, and a spin-symmetry breaking state may appear
according to the Stoner criterion. Whether or not this happens depends
on the density of states, which at least in the case of two dimensions
remains finite even down to $ n = 0 $.  Numerical simulations indicate
the tendency toward spin ordering is grossly exaggerated in mean field
theory \cite{p127491}. In any case a deeper discussion of this point is
not within the scope of the present article, which is to investigate
the analog of a Fermi liquid near half filling.

The Mott singlet  $ \ket{ \Phi_s (r) } = \Deltasd{r} \ket{0} $  at site $ r $ is created 
from the vacuum by the operator
$ \Deltasd{r} $ given by 
\begin{equation}
\Deltasd{r} = \frac{1}{2 \sqrt{2}} \sum_s e^{ i \pi (s + \frac{1}{2}) } \cdop{s}{r} \cdop{-s}{r} \;.
\end{equation}
A natural first attempt to understand the system for small hole doping
is to try the same method as for the nearly half filled
spin $ 1/2 $ Hubbard model, {\em i.e.}, to make a particle-hole transformation 
$ \chatd{s}{r} = \cop{-s}{r} $, where $ \chatd{s}{r} $ and $ \chat{s}{r} $
are new local fermionic creation and destruction operators.  In contrast
to the case of the filled spin 1/2 Hubbard model however, this canonical
transformation fails a number of criteria if the Mott singlet is 
to act as a vacuum for the new operators. In particular,
we see that
\begin{eqnarray} 
    \chat{s} {r} \ket{\Phi_s (r) } & \neq & 0%
    \label{problem1} \\
    \bra{ \Phi_s (r) } \chat{s} {r} \chatd{s} {r} \ket{\Phi_s (r) } & \ne & 1%
    \label{problem2} \ \ .
\end{eqnarray}

The first inequality is perhaps not so severe; after all the operator $
\chat{s}{r} $, being fundamentally a fermion creation operator 
creates a state with three fermions on site $ r $, and we
could argue that we could ignore this problem by suitably projecting
the states on those with two or less particles. This is in fact the
approach usually taken in attempts to perturbatively construct a new vacuum
at $ n = 2 $.  The second inequality is much worse; it is a consequence of the
singlet being  entangled, {\em  i.e.} it  cannot be written as a
product state in any one-particle basis.
As a result, the putative creation operator does not generate
a normalized state from the vacuum.  The entanglement property implies that
the destruction operator on the created state does not recreate the ground 
state---not even in the two particle space since it projects into the $ S=2 $ 
two-particle states. 

\section{Canonical transformation to the Mott singlet vacuum}
We now show how to systematically construct creation and annihilation
operators that  have the correct local properties.  We seek 
 a canonical transformation that fulfills the following
criteria: (a) it maps $ \ket{\Phi_s(r)} $ to $ \ket{0} $, and (b) it maps
the singly charged states to themselves. Due to our choice of interaction
parameters, where $ J \gt 2 U $, we also expect that the state  composed of two
holes will play the same role in the hole doped Mott insulator near
$ n = 2 $,  as the doubly charged singlet does for small filling. The
canonical transformation that we desire therefore has the property
that it interchanges the Mott singlet and the vacuum, leaving all
other states invariant.  The $ S = 2 $ doubly charged states will have
the same charge as the Mott singlet. However, due to the constraints
imposed by a canonical transformation utilizing only the spin $ 3/2 $
fermion operators,  these doubly charged states must be obtained by
two applications of the new creation operator, while the singlet created
by another double combination of these creation operators forms a
state with relative charge minus two.  These considerations force the
canonical transformation to be charge non-conserving.

The canonical transformation that accomplishes this and similar mappings can be systematically obtained
through the method in Ref. \cite{mele}. However, we can obtain it without
much formalism through the following argument.  Our desired operator is
``almost'' $ \Delta^{\dagger}_r $. The problem is  that this operator
generates unwanted side effects in the $ n= 1 $ and $ n = 2 $ particle subspaces 
by mapping these to new states with $ n = 3 $ and $ n = 4 $.  We
can get rid of these unwanted overlaps by using a projection operator, and  therefore
define $ \Qd{r} $ by $ \Qd{r} = \Deltasd{r}(1-n)(1-\frac{n}{2}) $. It is
straightforward to check that $ \Qd{r} \ket{0} = \ket{\Phi_s(r)} $,
that $   \Q{r} \ket{ \Phi_s(r) } = \ket{0} $,  and that $ \Q{r} $ and
$ \Qd{r} $ annihilate all other states. A canonical transformation that rotates the 
states $\ket{0}$ and $\ket{ \Phi_s(r) } $ at each site $r$ into each other without affecting the other states is
 provided by the unitary operator
\begin{eqnarray} \label{eq:unitary}
U(\tilde{\phi_r} )&\equiv&e^{iG(\tilde{\phi_r} )}=\prod_r U_r(\tilde{\phi}_r)
\nonumber \\
U_r(\tilde{\phi}_r) &\equiv & e^{iG_r(\tilde{\phi_r} )}=  e^{ i  ( \tilde{ \phi}_r  \Qd{r} + {\tilde{\phi}}^{\star}_r \Q{r} ) } 
\end{eqnarray}
where $\tilde{\phi}_r \equiv \phi_r e^{i\chi_r}$ ($\phi_r$ and $\chi_r$ are real). On  $\ket{0}$ and $\ket{ \Phi_s(r) } $
the transformation becomes
\begin{eqnarray} \label{eq:unitary2}
U_r(\tilde{\phi}_r)|_{0,\Phi} =\cos\phi_r + i \sin\phi_r \,  (e^{i\chi_r}\Qd{r} + e^{-i \chi_r}\Q{r}) \ \ ,
\end{eqnarray}
whereas it is unity on all other states.
Choosing $\phi_r = \pi/2$, for all $r$, we obtain the canonical transformation that fulfills our 
criteria, {\em i.e.} it interchanges the empty state and the Mott state at each site without 
affecting the other states.

Applying the unitary transformation Eq. \ref{eq:unitary}, with $\phi_r = \pi/2$, the true vacuum state, $\ket{0}$, 
is mapped onto the Mott insulator at half-filling $\ket{\Phi_s}$.  The phase factors $e^{ i \chi_r }$
enter this state only as an overall phase $\sum_r \chi_r$, and can 
be neglected. In general, however, it is obvious from Eq. \ref{eq:unitary} that the  unitary transformation gives  a 
state where the phase factors 
enter  in a non-trivial way. In particular, this is the case for the slightly doped Mott insulator, which 
we will be interested in below. This will be mapped onto a state near the true vacuum state, which 
can then be analyzed with standard methods.

Note that the  phase factors, $e^{ i \chi_r }$, are crucial to 
retain local gauge symmetry in the same way as the complex
phases introduced into the Bogoliubov transformations are necessary
to restore gauge invariance in BCS theory. 

\section{A variational ansatz}
We now turn to a systematic variational analysis of the slightly doped Mott 
insulator using the canonical transformation in 
Eq. \ref{eq:unitary}, and use computer algebra to
handle the complicated fermion polynomials that occur.
In analogy with ordinary Fermi liquid theory, as well as the BCS theory of 
superconductivity, we will then search for a variational state with 
particle number given by $ n = 2 - \delta $ that is obtained from the vacuum by 
a canonical transformation $ e^{ i {\cal G}_{u} }$  depending on a set
of parameters $ {u}$.  We define the functions $ E(u) $  and $ N(u) $ 
by
\begin{eqnarray}
E(u)  & =  &  \bra{0} e^{ -i {\cal G}_{u} } H e^{ i {\cal G}_{u} } \ket{0} \\
N(u) & = & \bra{0} e^{ -i {\cal G}_{u} } N e^{ i {\cal G}_{u} } \ket{0} \ \ .
\end{eqnarray}
The values $ \{u\}$ which minimize $ E(u) $ define our variational ground
state 
$ e^{ i {\cal G}_{u} } \ket{0} $  with particle
number $ N(u) $.  We have seen that for $ n =  2 $, the transformation 
 ${\cal G}_{u} $ is simply $G({\frac \pi 2} e^{i\chi _r}) \equiv G_0 (\chi_r)$ 
given by Eq. \ref{eq:unitary} . We therefore expect that
near the Mott insulator, the relevant transformation will be given by
a further transformation close to the identity. We therefore make the
ansatz $ e^{i {\cal G}_{u} } = e^{ i G_0 (\chi_r) } e^{ i {\cal G}'_{u} } \equiv U_0 U'$.

Note that  since we can continuously rotate the Mott state
at $ n = 2 $ to the true vacuum by letting $ \phi  $ go from $ \pi/2
$ to zero,  we can generate a Mott singlet 
on a site either by having $U'_r =1$ and $ \phi_r = \pi/2 $, or by having 
$U'_r =U_r(\pi/2)$ and $ \phi_r = 0 $. 
In general, we can make a coherent superposition of empty 
and doubly occupied singlet sites, both by
letting $ \phi $ vary and by adding an onsite s-wave order parameter.
As could be expected, this indeterminacy leads to a numerical 
instability in the variational equations which we resolve 
by simply taking $ \phi_r = \pi/2 $ for all $r$, and not
further exploiting these variational parameters.

In order to construct an ansatz for $ e^{ i {\cal G}'_{u} } $ we first work
out $   e^{ -i G_0 (\chi_r) } H  e^{ i G_0 (\chi_r) }  $. This operator
is obtained by replacing each occurence of the fermion operator
$ \cdop{r}{s} $ by
$  e^{ -i G_0 (\chi_r) } \cdop{r}{s} e^{ i G_0 (\chi_r) }$
and similarly for $ \cop{s}{r} $.
This expression is   complicated, but it can nonetheless be worked out exactly
in terms of polynomials of $ \cop{s}{r} $ and $ \cdop{s}{r} $, since the fermion 
algebra at a site is closed. 
The exact expression, written
here for reference only, is given by
\begin{eqnarray}
\cdop{s}{r} \rightarrow & 
         c^{\dagger}_{s,r}
    \left( \left(
         \sups{\Delta}{\dagger}   e^{-2 i \chi}\, \left( 1 - n \right)  - 
         e^{2 i\chi}\,\Delta \right)  + 
       \left( \frac{:\sups{S}{2}:}{3} + n + \frac{:\sups{n}{3}:}{6} \right) 
      \right)  + \nonumber  \\
& (-1)^{( s + \frac{1}{2} )} e^{2 i \chi} {2}^{-\frac{1}{2} } \, 
    \left( - 1 + 
       \, e^{-2 i \chi} \sups{\Delta}{\dagger} + 
       \left( \frac{:\sups{S}{2}:}{3} + n + \frac{:\sups{n}{2}:}{4} \right) 
       \right)   c^{\phdagger}_{-s,r}  \ \ ,
\end{eqnarray}
where the subscripts are dropped on the right hand side. The notation $ :O: $ is used to 
indicate a normal ordered operator, {\em i.e.}  strings of  fermion operators
where all creation operators are anticommuted to the left and annihilation operators
to the right taking only into account the sign of the permutation. 
In this case, $ :n^2: = n^2  - n $ and $  :S^2: = S^2 - 15n/4  $.

The onsite interaction is zero in the vacuum and two particle singlet subspace. 
Since these are the only two states affected by the canonical transformation, this
interaction remains invariant, while the chemical potential
transforms according to
\begin{eqnarray} \label{eq:exactn}
n  \rightarrow &  2   -  \left(  n  -  
	\frac{5\,:\sups{n}{2}: }{4} -
	\frac{\,:\sups{S}{2}: }{3}  +
	\frac{:\sups{n}{3}: }{6} \right) \ \  .
\end{eqnarray}

Anticipating a mean field calculation under the assumption of 
no spontaneously broken global symmetries, we do a Wick decomposition of the 
onsite term, and calculate the expectation value according to
\begin{equation} \label{eq:onsitemf}
\expectation{ U n ( n - 2 ) \; +  \;  J \, P^2_r } =  - \nh U  \; + \;
 \nh^2\,\left( \frac{5\,J}{8} + \frac{3\,U}{4} \right)  + 
   2\,{\Deltah{}}^2\,U \ \ ,
\end{equation}
where a hat indicates the expectation value of an operator
composed of ordinary fermion operators evaluated in the state $U^\prime \ket{0}$ near
the physical vacuum. Similarly, the expectation values for the density and s-wave order parameter 
$ \Deltasd{r} $ become exactly
\begin{eqnarray}
\expectation{ \Delta^{\dagger}}   & = & 
	  \frac{(\Deltah{}^*)^2 }{2}  e^{ - i \,\chi} - 
   \Deltah{}\,e^{2\,i \,\chi}\,
      \left( 1 + \frac{| \Deltah{}|^2}{2}  -  \frac{\nh}{2}+ \frac{\nh^2}{16} \right)  - 
   {\Deltah{}^*\,\left( | \Deltah{}|^2 - \frac{\nh}{2} + \frac{ \nh^2 }{8}  \right) } \\
\expectation{ n} & = &  \nh \; + \;   2 \,
   \left( 1 - \frac{\nh}{4}\right)^2 \;  
	\left( 1 -  \frac{\nh}{2} \right)  \;  - \; \frac{ \nh \, | \Deltah{r} |^2 }{2} 
	\label{eq:nopmf} \ \ .
\end{eqnarray}
We now derive a similar expansion of the hopping operator. 
In this case the expressions become a terrible mess---far to complicated to write down
in their entirety. It is however possible to construct a systematic expansion in the number of
fermion operators, which is appropriate for small doping. To this respect, we
 take the entire expression, and rewrite
it exactly as a sum of normal ordered terms. We then truncate this expression
at fourth order in fermion operators and keep expectation values of on-site and nearest neighbor 
s-wave pairing amplitudes, ordinary hopping and density operators.  Defining
\begin{eqnarray} \label{eq:deltadef}
\Delta^{\dagger}_{r,\rp} =  &
	\sum_s &  \frac{ (-1)^{ (s + \frac{1}{2}) }}{2 \sqrt{2}} \cdop{s}{r} \cdop{-s}{\rp} \\
\hop_{r,\rp} = &  \sum_s  & \cdop{s}{r} \cop{s}{ \rp }  \label{eq:hopdef} \ \ ,
\end{eqnarray}
we find that the hopping operator, truncated
to fourth order in fermion operators, becomes
\begin{eqnarray} \label{eq:hoptrunc}
\expectation{ h_{r,\rp}}  & =  
 - \frac{5t}{4} \; \nh \;  ( e^{2 \, i\,\chi_{r}}\,\Deltah{r,\rp} + e^{- 2 \, i \,\chi_{\rp} }  \Deltah{r,\rp}^* ) 
 +\frac{t}{2} \;  e^{2 \, i\,( \chi_{r} - \chi_{\rp}) }\, \left( 1 - \nh \right)  \hopi{\rp,r}  -   \nonumber \\
 %& 	\left( e^{ -i( \chi_r + \chi_{\rp} ) } \Deltahd{r} - 
&    t \left( e^{ - 2 i \chi_r  } \Deltah{r}^* - 
	     e^{  2 i \chi_r  } \Deltah{r} \right) e^{-2 i \chi_{\rp} } \Deltah{r,\rp}^* +
   t   \left( 
	     e^{ -  2 i\chi_{\rp} } \Deltah{\rp}^* -  
	e^{ 2  i \chi_{\rp} } \Deltah{\rp}
	\right) e^{2 i \chi_{r} } \Deltah{r,\rp} + ...
\end{eqnarray}
Decomposing the   s-wave order parameters as,
\begin{eqnarray} \label{eq:decom}
  \Deltah{r,\rp} &=&  e^{-i\eta_{r,r'} }   \tilde  \Delta_{r,r'}   \\ 
       \Deltah{r}& = & e^{-i\eta_{r} }   \tilde  \Delta_{r}     \nonumber   \, ,
 \end{eqnarray}
where $\tilde  \Delta_{r,r'} $ is real,  we obtain, 
\begin{eqnarray}\label{eq:hopmf}
		\expectation{h_{r,r'}} =  -t e^{i(\chi_r - \chi_{r'} )} &[ \frac 5 4 \hat n (e^{i(\chi_r + \chi_{r'} -\eta_{r,r'} )  } 
\tilde\Delta_{r,r'} + {\rm h.c.}  )  -  \half (1-\hat n)  e^{i(\chi_r - \chi_{r'} )}  \hat h_{r,r'}  \\
          &+   \left\{ (e^{2i(\chi_r - \eta_r)}\tilde\Delta_r -  {\rm h.c.}   ) (e^{i(\chi_r + \chi_{r'} -\eta_{r,r'} )  } 
          \tilde\Delta_{r,r'} + {\rm h.c.}               \right\}     ] + ...  \, . \nonumber
\end{eqnarray}
Written in this way, the invariance under local gauge transformations is manifest (of course
provided that the original hopping term is supplemented with the usual electromagnetic
phase factor $\exp (i\int_r^{r'} d\vec r\,'' \cdot \vec A(\vec r\,''))$).

Note that the combinations $\chi_r + \chi_{r'} -\eta_{r,r'} $ and $\chi_r - \eta_r$ are gauge invariant, 
so it is physically meaningful to fix them  to some 
value -- this corresponds to locking the phase of the order parameter to those of the on-site s-wave density, 
{\em i.e.} the phase of the local Mott singlets. 
Similarly, by virtue of Eq. \ref{eq:hopdef}, the combination $e^{i(\chi_r - \chi_{r'} )}  \hat h_{r,r'} $ is also 
gauge invariant. In the following, we shall set $\chi_r + \chi_{r'} -\eta_{r,r'}  = \chi_r - \eta_r = 0$.

The full set of variational parameters are now  $\chi_r$, which characterize the transformation $U_0$, together 
with the parameters $\{ u\}$, used to characterize $U'=e^{i{\cal G}'_u}$.  In the subsequent mean field calculation, these 
will be $\hat n$, $\tilde\Delta_{r,r'} $, $\tilde\Delta_r$ and $h_{r,r'}$, where $\tilde\Delta_r$ is the amplitude of 
the local s-wave order parameter.  

It is now clear that in order to evaluate the expression in Eq. \ref{eq:hopmf}
we must have some information about the phases $\chi_r$. In the undoped
Mott state, they only contribute to an overall phase, and  are thus
completely random. If this would be true also in the doped state,
both terms in Eq. \ref{eq:hopmf} would average to zero because of the phase
factor $e^{i(\chi_r - \chi_{r'} )}$. In the doped state, however, it is
reasonable to assume that some short range correlation is generated
among the phases $\chi_r$. This would correspond to having
\begin{eqnarray}\label{eq:phaseav}
\expectation{ e^{i\chi_r } e^{-i\chi_{r'} } }= f(\vec r - \vec{r'}) \, .
\end{eqnarray}
Really, this should be shown by including the phases $\chi_r$ in the variational calculation, but this is
technically very hard to do, so we shall instead simply assume Eq. \ref{eq:phaseav} to hold and set
 $f(\vec r - \vec{r'}) = 1$ for nearest neighbors. Under this assumption, Eq. \ref{eq:hoptrunc}
 simplifies to,
\begin{eqnarray} \label{eq:hopmf2}
\expectation{ h_{r\rp} }  = & 
\frac{t}{2} \; \hat{h}_{r,\rp}\,\left( 1 - \nh  \right)  -
   \frac{5 \, t }{2} \; \Deltat{r,\rp}\,\nh  + ... - \frac{t}{16} \Deltat{r} \hat{h} \nh^2 \ \ ,
\end{eqnarray}
where for future reference we have included the lowest higher order term which
couples linearly to $ \Deltat{r}$.
We therefore find that the Hamiltonian expectation value to be minimized
is given by  the sum of Eq. \ref{eq:onsitemf} and Eq. \ref{eq:hopmf2}, subject to
total particle number given by Eq. \ref{eq:nopmf}.
This effective Hamiltonian looks very much
like an ordinary BCS Hamiltonian, corresponding to Eq. \ref{eq:hamiltonian} but 
with one dramatic difference, namely the presence of  a term
proportional to $ \Deltat{r,\rp}\,\nh $, as well as a
higher order term which couples linearly to the  s-wave pairing operator. 

\section{Mean field analysis}
The mean field Hamiltonian can be analyzed by several equivalent methods. 
In the spirit of what was just developed, we could {\em e.g.} 
make a Bogoliubov-Valatin canonical transformation and minimize the
energy of the retransformed vacuum. This variational procedure would 
precisely correspond to the canonical transformation $  e^{ i {\cal G}'_{u} } $ 
alluded to earlier.  The method we actually use generalizes easily 
to finite temperatures and arbitrary large number of terms in the
polynomial expansion of the mean field Hamiltonian.  It  uses
that the density matrix $ \rho = e^{ \beta ( \Omega(T,\mu)  -  ( H - \mu N )  )  } $ minimizes the free energy
$ F = \langle H - \mu N \rangle_{\rho} - \,  k T \langle S\rangle_{\rho} $ for all values of $ \rho $ 
($e^{-\beta\Omega}$ is the partition function and $S$ the entropy). We can then
take $ \rho $ to be the exponential of an expression linear in
$ \nh_r, \Deltah{r\rp}, \Deltah{r} $ and $ h_{r,\rp} $ , where the prefactors are 
varied to minimize $ F $.   This method yields the ordinary BCS theory 
when applied to a Hamiltonian of the form in Eq.  \ref{eq:hamiltonian} and
gives a more complicated self-consistent calculation when more terms are kept.

We have performed the mean field analysis numerically, both using the truncated
expressions given explicitly above and the full mean field theory
containing polynomials to seventh order. Since we construct an effective Hamiltonian,
we define the hatted operators whose expectation values give the values $ \Deltahd{r} $ in
Eq. \ref{eq:hoptrunc}.  The corresponding operators $ \Deltahd{r}, \Deltahd{r,\rp} , \hat{h} $ 
are therefore formally $ \Deltahd{r} =  e^{ -i {\cal G}_{u} }  \Delta^{\dagger}_{r} e^{ i {\cal G}_{u} } $ {\em etc.}   but in the calculation this involves simply reinterpreting 
the original operator in Eq. \ref{eq:deltadef} in terms of quasiparticle fermion operators. For the density matrix
$\rho \propto e^{ -\beta H_{eff} } $ we choose the Hamiltonian $H=H_{eff}$ as 
\begin{equation}
\label{eq:effham}
H_{eff} = \sum_{r,\rp}  ( t' \, h_{r\rp}  +  \,2^{\frac{3}{2}} \gamma  \; \Delta^{\dagger} _{r\rp} ) -
	  \sum_{r}      ( \mu' n_r   +   \,2^{\frac{3}{2}}   \gamma_2 \; \Delta^{\dagger}_{r}  ) \;  + \;  ( CC )
\end{equation}

\renewcommand{\dmu}{{\delta_\mu}}
\newcommand{\dg}{{\delta_\gamma}}
\newcommand{\bev}[1]{  \widetilde{ #1  }  }
\newcommand{\dimension} {D}

Our approximation is reasonable for small doping, and we confine
the mean field analysis to this regime.
We let $ \epsilon_k = ( 2 - \cos{k_x} - \cos{k_y}   ) $ in two dimensions with
a similar expression for $ \dimension = 3 $. With $\mu' = Dt' + \delta_\mu$ and
$\gamma_2 = D\gamma + \delta_\gamma$ we define 
$ e_k =    t' \, \epsilon_k    + \delta_\mu  $, 
$ d_k =    \gamma  \epsilon_k   +  \delta_\gamma $ and
$ E_k = \sqrt{ e_k^2 + d_k^2 } $.
The energy gap $ \Delta $ is the minimum in $ E_k $, and it is easy to verify that 
for small doping an excellent approximation is given by 
$ \Delta =   \frac{| \gamma_\dmu - \dg t' \,|  }{ \sqrt{  \gamma^2 + (t')^2} }$. 
We define the momentum space sums as $ \bev{ f(k) } = \sum_k \frac{ |t'| f(k) }{ E_k} $.
Taking into account that there are four spin values  we find 
the following  expression for the doping $  \delta \approx  \nh$,
where expectation values of operators are dropped when the context is clear,
\begin{eqnarray} \label{eq:anh}
 \nh  =  4 \, \times  \,  \frac{1}{2} {\displaystyle \int }
 \left( 1 - \frac{ e_k }E_k  \right) \; d^2 k  &  = 2 - 2 \, \bev{ e_k }/|t'| 
\end{eqnarray}
The expressions for $t' $ and $ \gamma$  can
be read off to lowest order from Eq. \ref{eq:hopmf2}
\begin{eqnarray}
 t' = &  - \frac{1}{2} t ( 1 - \nh )   \\
\gamma  =  &  -  \frac{ 5  \nh t }{  4 \sqrt{2} }    \label{eq:gamma} \, ,
\end{eqnarray}
while the definition of the on-site s-wave order parameter can be read off from
Eq. \ref{eq:onsitemf}
\begin{eqnarray}
 \dimension \gamma - \dg  & =   2 \sqrt{2} U  \Delta_r   \, . 
\end{eqnarray}
In writing Eq. \ref{eq:effham}, we neglected the non-local repulsive interactions of
type $n_r n_{r'}$ that will certainly be present in any realistic model with screened 
Coulomb interaction. Assuming a nearest neighbor term, $U_1 n_r n_{r'}$ and using
the identity $\langle n_r n_{r'}\rangle = 1 + \frac 1 2 \Deltah{r,r'} \Deltahd{r,r'} - \frac 1 4 
h_{r,r'}h_{r',r} - 2(\hat n_r + n_{r'} )+  \frac 5 4( n_r^2  + n_{r'}^2)  +  n_r n_{r'}  $, Eq. 
\ref{eq:gamma} would change to
\begin{eqnarray}
\label{eq:gamma2}
\gamma  =  - \frac{ 5 \nh t }{  4 \sqrt{2}  }  +  \sqrt{2}  \, U_1 \, \Deltatd{r,r'} \
\end{eqnarray}
Below we argue that this would not qualitatively change our conclusions. 

As usual, self-consistency implies a gap equation which here reads,
\begin{eqnarray}
  \Delta_r = -  \frac{1 }{\sqrt 2t'}  \bev{d_k}  \, .
\end{eqnarray} 
After expanding to lowest order in $ \delta $ and doing some some algebra this can be recast as
\begin{equation} \label{eq:selfc}
\frac{t}{4\,U} \left( \frac{\Delta }{\sqrt{2} t  } - \frac{5 \dimension \delta}{8} \right) = 
 \frac{5 \delta}{8} - \frac{\Delta}{t \, \sqrt{2} } \bev{1} \, ,
\end{equation}
which is a closed equation that can be used to find  the physical gap $ \Delta $  as a function
of $ \delta $.  

Using the self-consistent equations, we can find expressions for
the extended  and onsite s-wave pairing amplitudes:
\begin{eqnarray}
\Deltat{r,\rp} & \approx&    \left( \frac{5 \dimension  \delta}{8} -  \frac{\Delta }{\sqrt{2} t  }  \right) \\
\Deltat{r} & \approx &  - \frac{t}{2\,U} \Deltah{r\rp}  \, . \nonumber
\end{eqnarray}
Thus, not surprisingly we find that the onsite s-wave component  is reduced
by a factor $ t/U $ relative to the extended component.  We note that the particular
combination corresponding to $ \Delta_{r\rp} $ occurs in the left hand side of
Eq. \ref{eq:selfc} which is consistent with the onsite s-wave component  vanishing
while the nearest neighbor extended pairing remains finite in the limit $ U \rightarrow \infty $.
We can now understand what would be the qualitative effect  of adding extra repulsive
interactions corresponding to the redefinition Eq. \ref{eq:gamma2} of the variational parameter 
$\gamma$. For large $U_1$, $\gamma$ will effectively be put to zero corresponding to 
$\Deltat{r,r'} \sim \frac {t'} {U_1} \hat n$, rather than $\Deltat{r,r'} \sim  \delta$. We see that the 
scale of $\Deltat{r,r'} $    changes but it is still non-zero for arbitrary small doping.

\subsection{ Asymptotic behavior of the gap in $\dimension = 2 $ }
In two dimensions, the  expression Eq. \ref{eq:anh} can be approximated by
\begin{eqnarray}  \nh  = \approx  
        \frac{1}{\pi \, |t'| }  ( \delta_\mu + \sqrt{ \delta_\mu^2 + \Delta^2 } ) \, .
\label{eq:anh2}
\end{eqnarray}
The expression for $  \bev{1} $ is logarithmically singular but can
be approximated by \begin{equation} \label{eq:logsing}
 \bev{1} \, \approx  \frac{-1}{2 \pi  } \ln{\left( \frac{ - \dmu \, + \, 
        \sqrt{ \delta_\mu^2 + \Delta^2 } }{32 \, |t'| } \right) }   = 
        \frac{1}{2 \pi} \ln \left( \frac{ 32 \nh \pi (t')^2 }{\Delta^2 } \right)  \ ,
\end{equation}
where Eq. \ref{eq:anh2} was used.
In spite of the logarithmic singularity, the self-consistent equation Eq. \ref{eq:selfc} can
be solved in closed form. Defining $ q = \Delta/( t \delta ) $  
the (inverse) equation is
\begin{equation} \label{eq:numapprox}
 \delta  \approx  \frac{8 \pi}{q^2} \,e^{\frac{\pi \,\left( 4\,q - 
            5\,{\sqrt{2}}\,\left( 1 + 2\,U/t \right)  \right) }{8\,q\,U/t}}  \, .
\end{equation}

By plotting the pairs $ ( \delta, q \delta )  $ according to the above formula as 
a function of $  q $, we find the gap as a function of $ \delta $, shown for 
values of $ U=(\infty,10,5,1) $ in Fig. 1.   We can see an almost linear
behavior of the gap as a function of doping that is quite insensitive
to the value of $ U $. For all values of $ \delta $ and $ U $
the approximation $ \Delta \approx  t \delta $ is a surprisingly good
approximation. A comparision with the numerical solution of the self-consistent
equations is shown in Fig. 2.

\subsection{ Asymptotic solution for $ \dimension = 3 $  and small doping}
In the case $ \dimension  = 3 $, the vanishing density of states near
$ k = 0 $ makes the integral $ \bev{1} $ converge.
In this case the self-consistent equation
is Eq. \ref{eq:selfc}  with $  \dimension = 3 $, and constant $ \bev{1}$. However,
due to the vanishing density of states, even relatively small values 
of density  lead to quite substantial values of $ \mu $ and $\delta $ 
so the asymptotic value of this equation is far from being reached
even for doping as  low as $ .01 $. The corrections to $ \bev{1} $ 
are rather slowly varying, so the linear dependence of the gap upon doping
is obtained for the $ \dimension = 3 $ case as well, as shown
in Fig. 2.

\section{Effective theory for small $\delta$ }

After applying a transformation that rigorously preserves the full 
Fock space, we have obtained a
non-Fermi liquid behavior for the doped spin 3/2 Mott insulator by using
standard mean field theory methods. 
Here we contrast this to an effective 
theory for the Hamiltonian in Eq. \ref{eq:hamiltonian}, derived in the limit of  small doping $\delta \ll 1$ 
and small hopping $t \ll U,J$, using the more conventional approach of projection 
on a low energy subspace. 
  Ignoring hopping, $t=0$,
this low energy sector consists of states where each site is occupied by either a Mott singlet 
$\ket{\Phi_s  }$, which we choose as the Fock vacuum, $\ket{\Phi_s  } \rightarrow \ket{0}$, or by  a 
single charge $\ket{s}={c}^\dagger_s \ket{0 }$ with spin 3/2 ($s=\pm 1/2, \, \pm 3/2$) -- all other 
states are separated from these by a gap of order $X \sim U,\, J$. Restricting to this low energy 
sector and including the hopping in perturbation theory gives to order $t^2/X,$  the effective Hamiltonian
\begin{eqnarray}\label{eq:effmodel}
H_{eff} =-\tilde t  \sum_{r,\delta,s} c^\dagger_{sr} c_{s,r+\delta} + 
\sum_{r,\delta, \alpha, \alpha^{\prime},\beta, \beta^{\prime} } {\tilde{J}}_{\alpha \alpha^{\prime}\beta\beta^{\prime} } 
c^\dagger _{\alpha,r} c_{\alpha^\prime r }c^\dagger_{ \beta r+\delta} c_{\beta^{\prime} ,r+\delta}
\end{eqnarray}
subject to the constraint  $\sum_{s} c^{\dagger}_{sr} c_{sr} \le 1$ 
($ {\tilde{J}}_{\alpha \alpha^{\prime}\beta\beta^{\prime} }$ are $ SU(2) $ scalars). This $t-J$-type  
model describes four species of fermions  with nearest neighbor hopping ($\tilde t  \sim t$), nearest neighbor
exchange  couplings
($\tilde J  \sim t^2/X$) and with the hard-core constraint that no two fermions occupy the same site. 
Second order perturbation theory guarantees a finite,
albeit weak, attraction which opens the possibility of having a
superconducting phase even for small doping. From this approach, however, we would expect
such a phase to be destroyed by a nearest neighbor repulsion that is normally
present in a realistic model. Thus, our previous mean field calculation is at odds with this approach. 
If the former turns out to be valid, it suggests that there are non-perturbative effects due to the 
hard-core constraints that are not easily accounted for in the conventional formulation. If on the
other hand the hard-core constraints are not very important and the naive picture of 
four different species of weakly interacting fermions is essentially correct, it would suggest
that our mean field treatment of the phase phase fluctuations  does not capture the correct
physics.

\section{Discussion and summary}
\subsection{The anomalous term $\Delta_{r,r'}n_r$. }
We see from Eq. \ref{eq:nopmf} that in order to have a nonzero doping
$ \delta = 2 - \langle n\rangle $,  we {\it must } have $ \nh \gt
0 $, in fact $ \nh \approx \delta $ to lowest order in $ \delta $.
Energetically we will also have $ \hat{h}  \ne 0 $ for finite $
\delta $.    According to Eq. \ref{eq:hopmf} the extended s-wave pairing
field $\Deltah{r,\rp}$ cannot vanish and in fact will be proportional
to doping. This in turn generates a (much smaller) on-site pairing
$\Deltah{r}$ through the self-consistent equations.  Note that this
pairing field can never completely vanish because of the linear coupling
to higher order terms.  At finite temperature the mean field theory will
presumably eventually break down via an $xy$-transition due to phase
fluctuations that  we have not taken into account. 
This has been discussed in a series of recent papers where the term
``gossamer superconductivity'' \cite{goss1,goss2,condmat0312573} has
been used to describe a similar scenario.

It is admittedly not easy to understand the physical origin of these new anomalous
terms of the type $\Delta_{r,r'} n_r$. 
On a technical level, they are  forced
by the fermion statistics which  constrains the form of the  canonical
transformation necessary to map a local Mott ground state to the vacuum.  In our
case, this transformation must be (a) nonlinear in fermion operators and
(b) charge  non-conserving.  Property (a) yields an effective interaction
from the hopping term near a charged ground state and property (b) makes
this interaction non-gauge invariant. 
Property (a) is a necessary consequence of mapping a locally entangled state 
to the vacuum and property (b) is a conseqence of mapping a charged state
to the vaccum which breaks gauge symmetry.  Very general arguments relying 
on long range phase coherence and a finite range gap function then predict that the
system should be a superconductor \cite{forster}. 
Our mean field calculation, which suggests a superconducting ground state, 
supports this picture, given our assumptions about phase coherence. 

We already pointed out the contradiction between  our main result and
what would be expected based on a conventional analysis 
of the type leading to  Eq. \ref{eq:effmodel}, but also stressed 
the difficulties 
related to the hard core constraints inherent in the latter approach. 
Here we should note that more elaborate schemes for dealing 
with these non-holonomic hard core constraints  face severe
difficulties related to phase fluctuations. For example, in the spin 1/2 Hubbard model
at half-filling, one can turn the no double occupancy constraint into
a holonomic gauge constraint by introducing spinons and holons. The resulting
phase depends crucially on the fluctuations in the related gauge fields. 
By working in the full Hilbert space, we avoid these difficulties, but 
nevertheless our conclusions are still 
dependent on certain
assumtions about phase coherence.  Without a more sophisticated analysis of the 
phase fluctuations, we cannot rule out that these will be important and
{\em e.g.} destroy the superconduting state at low doping.

\subsection{Range of validity and applicability}
We now assume that our analysis is correct at low doping, and discuss its range of validity
and applicability. 
At sufficiently large value of doping, the theory will yield a free
energy which is unfavorable compared to that of a doped $ n= 1 $ Mott
insulator. The mean field picture suggests there will be a coexistence
region where a slightly hole doped $ n= 2 $ Mott insulator will coexist
with a hole doped $ n = 1 $ Mott insulator.  The $ n = 1 $ Mott insulator
will presumably have some sort of magnetic order at low temperature
that breaks the large spin degeneracy of the uncorrelated odd filling
Mott insulator.  If a coexistence region really exists,  or whether an
intermediate phase which breaks translational invariance may exist,  is beyond
the scope of the present analysis.
Our calculation thus  makes assumptions about  $ U $ which leave
open the question if this behavior could really be seen in a physical
system. On the one hand, $ U $ must be large enough (and $ J $ even larger)
so that a Mott insulating state occurs at $ n = 2 $ and furthermore triply
occupied sites are effectively absent.  On the other hand, $ U $ must be
small enough so that the correlated state will have  lower energy than a mixed
state with an $ n = 1 $ Mott insulator and an $ n = 2 $  Mott insulator.

We should also ask how dependent our approximations are on 
our specific choice of parameters. 
In particular what would happen
if we had taken $ J <  0 $ in the spin $ 3/2 $ model? According to the
analysis in Ref. \cite{PRL8642}, the ground state at $ n=2 $ is than $
SO(5) $ invariant and we could choose for instance $ \cdop{\frac{3}{2}}{r}
\cdop{\frac{1}{2}}{r} |0\rangle $ as the Mott vacuum. However, this as
well as the other states with $ m \ne $ 0  are {\it not } entangled, and
the procedure discussed here results in a simple particle-hole
transformation of pairs of fermion operators which does not generate
any anomalous terms and the ordinary BCS-type analysis presented in these
calculations should be valid.  However, the state $ m = 0 $ is entangled, and
it can be checked by the methods of Ref. \cite{mele} that the canonical
transformation that maps this state to the vacuum will be nonlinear in
fermion operators as well as charge nonconserving causing an effective
interaction and superconductivity to appear at finite doping through a
similar mechanism discussed here for the Mott singlet. 

If we take the parameters as $ 0  <  J    <    2
U $,  which of course is more physically reasonable, the
energetically most favorable state constructed with two quasicharge
operators $ \chatd{s}{r}$ from the Mott singlet will no longer be a
singlet with relative charge minus two, but rather one of the $ l=
2 $ multiplets with zero relative charge.  The extended s-wave order
parameter will still appear since it is a direct consequence
of mapping the Mott singlet to the vacuum, but there will be a spin
order parameter introduced that takes the place of the local s-wave
superconducting order parameter. The presence of an additional order
parameter together with spin degeneracy makes this calculation more
difficult and it awaits further analysis.

Finally, it is  relevant  to ask whether the transformation used 
for the spin 3/2 case  could be applied to the  spin 1/2 systems. 
First consider the canonical transformation which maps between these the $n=0$ and 
$n=2$ states. This is
the ordinary particle-hole transformation which is not charge conserving.
However, the doubly occupied singlet is created by $ \cdop{\uparrow}{r}
\cdop{\downarrow}{r} |0\rangle$ and hence is factorizable in fermion
operators. The canonical transformation is therefore linear and no  new
interaction terms are introduced in the transformation.  The physical
properties of the system are symmetric under charge conjugation, which
is sufficient for the particle-hole transformation not to generate any
new behavior and the present analysis is uninteresting.  In the case of
the half-filled Hubbard model, the Mott ground state corresponds to one
electron per site. This cannot be mapped to the vacuum through a canonical
transformation without globally violating the Fermi anticommutation
relations\cite{mele}.

\subsection{Summary}

We have presented a new type of canonical transformation for
the half-filled spin $ 3/2 $ Hubbard model that maps the Mott insulator 
at half filling to the vacuum. 
This canonical transformation is straighforward to
generalize to multi-band Hubbard models with a local spin singlet Mott 
insulating ground state. 
At finite doping, a self-consistent mean field theory for such a system results in a
phase with long range phase coherence.  An order parameter that
is usually identified with extended s-wave superconductivity
appears and is proportional to doping.  The calculation appears to be
in contradiction to other methods of attacking these kinds of problems, 
and we pointed out the difficulties with both approaches. 
Our calculations bear a striking resemblance to the  ``gossamer superconductor''
scenario that has been recently introduced by Lauglin and coworkers.
Although we have only explored a specific half-filled spin $ 3/2 $ Hubbard model, we 
believe that our method could be useful for variety of similar models with 
locally entangled Mott insulating ground states.

\begin{figure}
\includegraphics[scale=1.0]{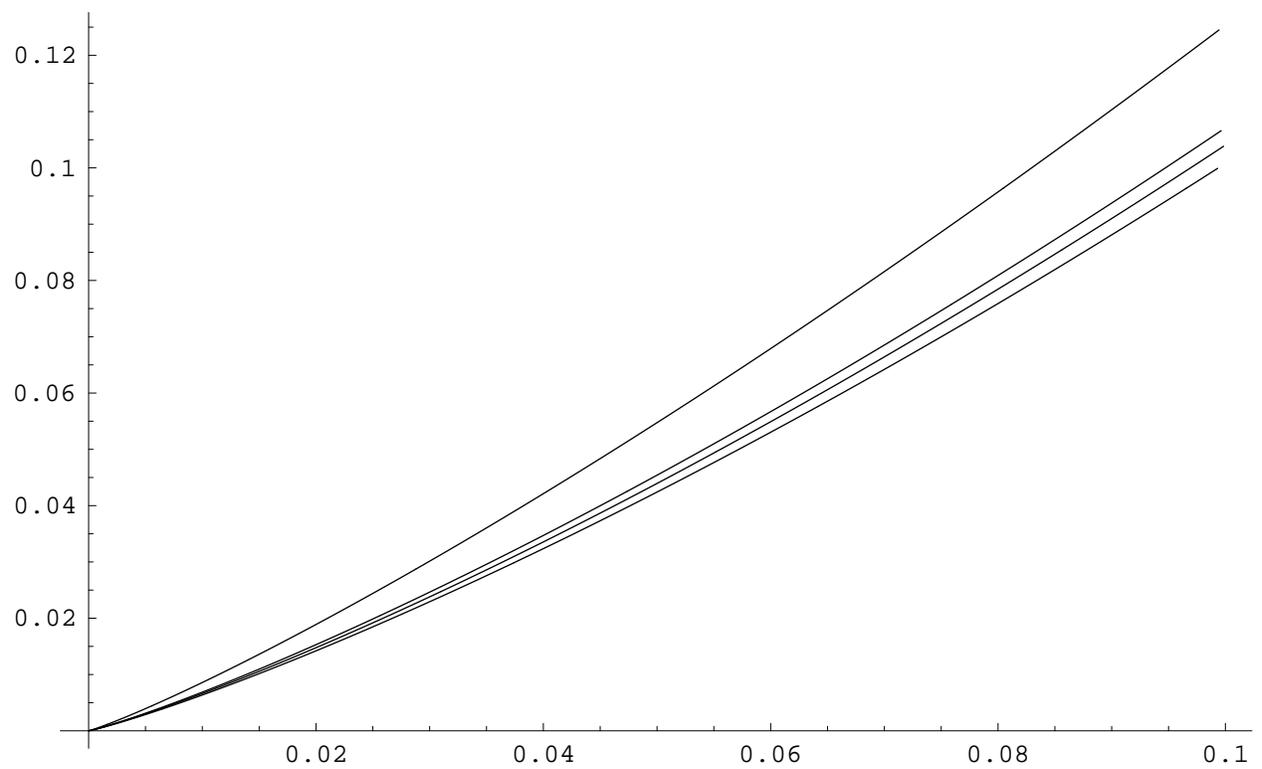}% Here is how to import EPS art
\caption{\label{fig:qvsdelta} 
Values of $ \Delta/t  $ as a function of doping $ \delta $
for $ U  = (\infty,10,5,1 ) $ for $ \delta  \le 0.1 $, with
the gap decreasing monotone with increasing $ U $ at given
$ \delta $.  }
\end{figure}

\begin{figure}
\includegraphics[scale=1.0]{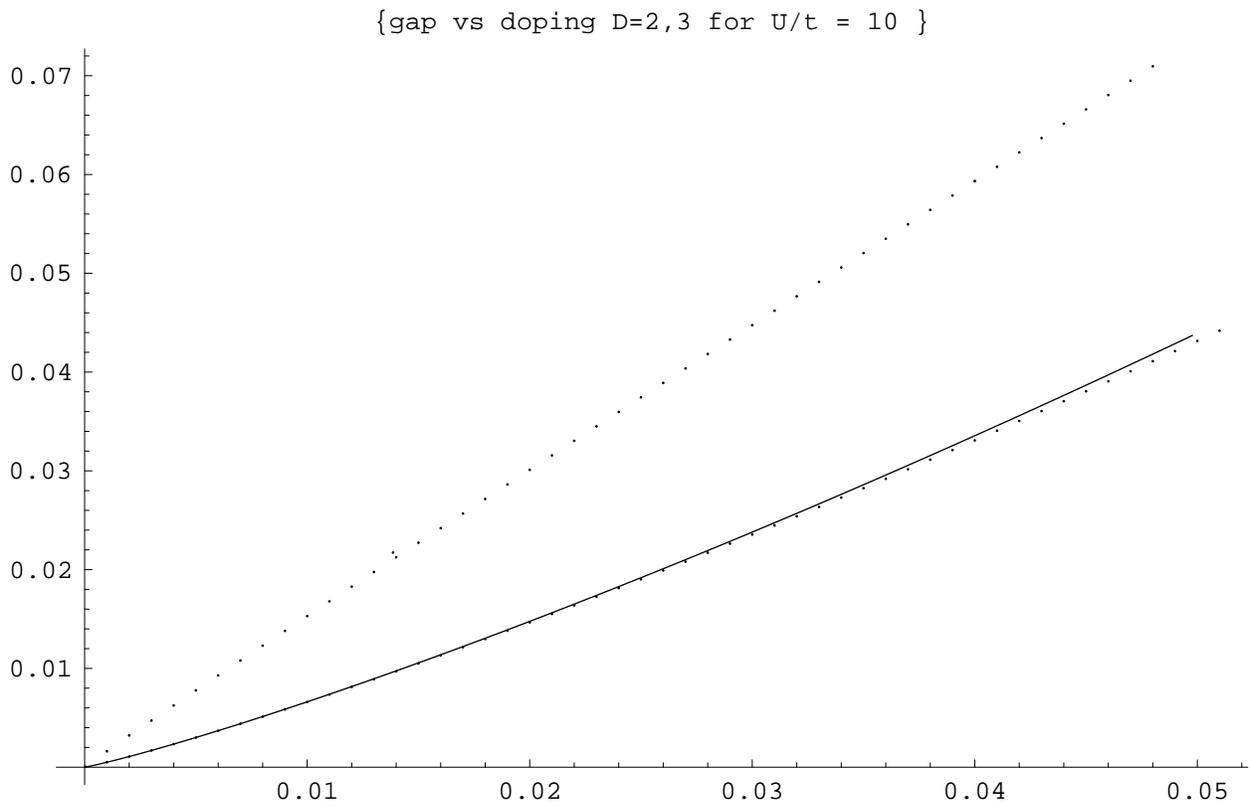}% Here is how to import EPS art
\caption{\label{fig:numdata} 
Values of $ \Delta/t  $ as a function of doping $ \delta $
for $  U/t = 10 $ for $ \delta  <  0.05 $. Th upper curve
is for $ D = 3 $ and the lower data points for $ D = 2 $.
The solid line is the fit to the $ 2 \, \dimension $ asymptotic
curve according to Eq. \ref{eq:numapprox}. The  $ 3 \, \dimension $ fits
with no visible error to the curve $ \Delta/t = 1.4848 \delta  $ corresponding
to $ \bev{1} = .6137 $.
}
\end{figure}

%\bibliography{gbvprl}

\end{document}